\documentclass[journal]{IEEEtran}

\ifCLASSINFOpdf
\else
   \usepackage[dvips]{graphicx}
\fi
\usepackage{url}
\usepackage{amsmath,amsfonts}
\usepackage{bm}
\usepackage{xcolor}
\usepackage{tabularx}
\usepackage{comment}
\usepackage{graphicx}
\usepackage{hyperref}

\begin{document}

\title{Efficient LOS-Sampled GNSS Direct Position Estimation: An Information-Loss CRB Analysis}
\author{Wei Gao, Rong Yang,~\IEEEmembership{Senior Member,~IEEE,} Jihong Huang, and Xingqun Zhan,~\IEEEmembership{Senior Member,~IEEE,}


\thanks{Supported by the National Natural Science Foundation of China (U2570201). AI-assisted tools were used for language polishing. Submitted to IEEE for possible publication. {\itshape (Corresponding author: Rong Yang.)}}
\thanks{Wei Gao, Rong Yang, and Xingqun Zhan are with the School of Aeronautics and Astronautics, Shanghai Jiao Tong University, Shanghai 200240, China (e-mail: gaowei1515@sjtu.edu.cn; xqzhan@sjtu.edu.cn; rongyang@sjtu.edu.cn).}
\thanks{Jihong Huang is with the Ann and H. J. Smead Department of Smead
Aerospace Engineering Sciences, University of Colorado Boulder, CO 80303, USA (e-mail: jihong.huang@colorado.edu).}}

\markboth{Journal of \LaTeX\ Class Files, Vol. 14, No. 8, August 2015}
{Shell \MakeLowercase{\textit{et al.}}: Bare Demo of IEEEtran.cls for IEEE Journals}
\maketitle

\begin{abstract}

Conventional Global Navigation Satellite System (GNSS) Direct Position Estimation (DPE) exploits raw intermediate-frequency (IF) data and provides a full-information Cramér-Rao Bound (CRB) benchmark, but its accumulated-correlation objective requires dense evaluations over a common Position, Velocity, and Time (PVT) search space. This paper proposes an efficient Line-of-Sight (LOS)-sampled DPE, where each satellite channel independently retains only PVT sample points aligned with its LOS direction. A residual-minimization estimator is formulated to resolve the mismatch between accumulated-correlation metrics and per-satellite LOS sampling. The Fisher information matrix (FIM) and information-loss CRB of LOS-sampled DPE are derived, quantifying the information loss determined by LOS sampling parameters. Theoretical analysis, Monte Carlo simulations, and real experiments show that proper LOS sampling approaches the full-information CRB and practical performance of conventional DPE, while reducing the number of correlation evaluations from exponential to linear growth.
\end{abstract}


\IEEEpeerreviewmaketitle

\section{Introduction}

\IEEEPARstart{D}{irect} Position Estimation (DPE) has attracted increasing attention in Global Navigation Satellite System (GNSS) receivers because of its improved robustness in challenging propagation environments, such as urban canyons and multipath-rich scenarios \cite{DPE Multireceiver}-\cite{DPE Urban test}. Unlike conventional two-step receivers, DPE estimates position, velocity, and time (PVT) directly from raw intermediate-frequency (IF) data and jointly exploits the signal-level information from multiple satellite channels \cite{DPE review}. Under this raw IF observation, the Cramer--Rao Bound (CRB) provides the full-information benchmark for attainable PVT accuracy of DPE receivers \cite{DPE Cramer–Rao Bound}-\cite{DPE CRB analytic}. However, practical DPE usually requires a high-dimensional search over the PVT space and dense correlation evaluations, making computational complexity a major obstacle to implementation.

\begin{figure}[ht]
    \centering
    \includegraphics[width=1\linewidth]{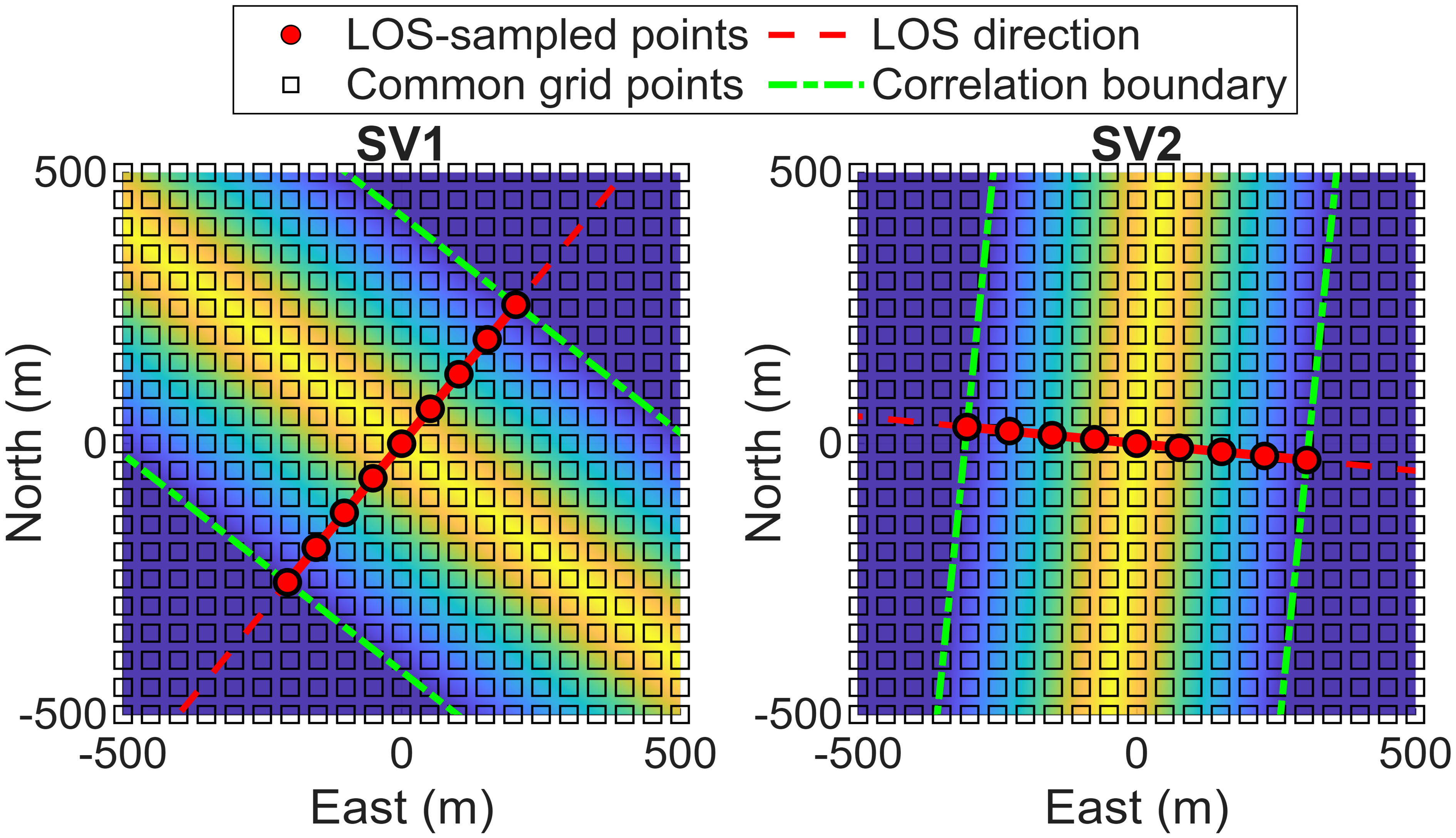}
    \caption{Illustration of conventional full-space dense correlation (black squares) and the sparse per-satellite LOS sampling (red dots) in the local position plane.}
    \label{fig:los_sampling_intro}
\end{figure}

Considerable effort has therefore been devoted to reducing the computational burden of DPE. Existing approaches include multi-resolution or grid space reduction \cite{DPE multi-resolution}-\cite{Search space reduction}, faster correlation computation through duty-cycling, parallelization, and multicorrelator interpolation \cite{Duty-Cycling}-\cite{DPE correlator interpolate}, and stochastic optimization methods such as particle filtering and swarm intelligence \cite{PF weighting DPE}-\cite{DPE swarm opt}. Although effective, these methods generally retain the conventional accumulated-correlation metric over a common PVT search space \cite{DPE MLE transform}. From the perspective of an individual satellite channel, however, many correlation values on these common grid points are weakly informative or locally redundant \cite{Collective detection}. This motivates a channel-dependent Line-of-Sight (LOS) sampling that preserves dominant PVT-related information with fewer correlation evaluations, as illustrated in Fig. \ref{fig:los_sampling_intro}.

This paper develops a LOS-sampled DPE framework and establishes its information-loss CRB. The contributions are threefold: 1) motivated by the fact that the correlation function varies most rapidly along the LOS direction in the PVT space \cite{Geometric DPE}, we convert this geometric property into an explicit per-channel sampling rule. Instead of evaluating a dense full-space correlation field over common PVT candidates, each satellite channel retains only a limited number of correlation samples along its own LOS direction. 2) we formulate the corresponding LOS-sampled DPE estimator based on these channel-dependent correlation measurements, thereby resolving the structural mismatch between conventional accumulated-correlation metric and per-satellite LOS sampling. 3) we derive the Fisher information matrix (FIM) for the LOS-sampled correlation measurements and obtain an information-loss CRB. Since LOS sampling keeps only a one-dimensional subset of the full PVT-space correlation field for each channel, part of the raw full information is inevitably lost. The proposed CRB explicitly quantifies this loss through a normalized information factor determined by the sampling step size and the number of samples. Theoretical analysis, Monte Carlo simulations, and real experiments show that proper LOS sampling can approach the full-information CRB benchmark and practical performance of conventional DPE, while reducing the number of correlation evaluations from exponential growth with the PVT dimension to linear growth with the number of LOS samples.

\section{The LOS sampling strategy for DPE}

\subsection{The measurement model of correlation values}

The PVT parameters estimated in DPE receivers are typically defined as \cite{DPE definition}
\begin{equation}
{\bm{\gamma }} = {\left[ {\begin{array}{*{20}{c}}
{{{\bf{p}}^T}}&{c\delta t}&{{{\bf{v}}^T}}&{c\delta \dot t}
\end{array}} \right]^T}
\label{eq:PVT_define}
\end{equation}
where ${\bf{p}} = {\left[ {\begin{array}{*{20}{c}}{{p_x}}&{{p_y}}&{{p_z}}\end{array}} \right]^T}$ and ${\bf{v}} = {\left[ {\begin{array}{*{20}{c}}{{v_x}}&{{v_y}}&{{v_z}}\end{array}} \right]^T}$ are the position and velocity of the receiver. The clock bias $c\delta t$ and the clock drift $c\delta{\dot t}$ are normalized by the speed of light $c$. For the $m$-th satellite channel, the complex correlation value at a channel-specific PVT sample point $\tilde{\bm{\gamma}}^m$ is modeled as

\begin{equation}
{\Re ^m}\left( {{\bm{\tilde \gamma }}^m} \right)  = h_{}^m\left( {{\bm{\tilde \gamma }^m},{\bm{\gamma }}} \right) + \nu _{}^m\left( {{\bm{\tilde \gamma }}^m} \right)
\label{eq:Correlation_model}
\end{equation}
where the superscript ‘$\sim$’ denotes quantities evaluated at PVT sample points.

The term $\nu ^m\left( {{\bm{\tilde \gamma }}^m} \right)$ in (\ref{eq:Correlation_model}) is the measurement noise. The expression $E\left[ {\nu ^m\left( {{{{\bm{\tilde \gamma }}}^m_i}} \right)\nu ^m\left( {{{{\bm{\tilde \gamma }}}^m_j}} \right)}^* \right]$ describes the noise covariance between different sample points ${\bm{\tilde \gamma}}^m_i$ and ${\bm{\tilde \gamma}}^m_j$.
\begin{equation}
E\left[ {\nu ^m\left( {{{{\bm{\tilde \gamma }}}^m_i}} \right)\nu ^m\left( {{{{\bm{\tilde \gamma }}}^m_j}} \right)}^* \right] \approx \frac{{{N_0}}}{{4{T_c}}}\Re _\tau ^m\left( {\nabla \tau _{i,j}^m} \right)\Re _f^m\left( {\nabla f_{i,j}^m} \right)
\label{eq:Spatial_coherence}
\end{equation}
where $T_{c}$ is the coherent integration time, and $N_0$ is the noise power spectral density. The difference terms $\nabla \tau _{i,j}^m = \tilde \tau _j^m - \tilde \tau _i^m$ and $\nabla f_{i,j}^m = \tilde f_{j}^m - \tilde f_{i}^m$ represent the differences of code-delay and Doppler between sample points ${\bm{\tilde \gamma}}^m_i$ and ${\bm{\tilde \gamma}}^m_j$. The correlation functions $\Re _\tau ^m\left( \cdot \right)$ and $\Re _f^m\left( \cdot \right)$ are given as
\begin{equation}
\Re _\tau ^m\left( {\Delta {\tau ^m}} \right) \approx \max \left( {0,1 - \left| {\Delta {\tau ^m}} \right|} \right)
\label{eq:CAF_tao}
\end{equation}
\begin{equation}
\Re _f^m\left( {\Delta f^m} \right) = \frac{{{\rm{sin}}\left( {\pi \Delta f^m{T_c}} \right)}}{{\pi \Delta f^m{T_c}}} \buildrel\textstyle.\over= {\rm{sinc}}\left( {\Delta f^m{T_c}} \right)
\label{eq:CAF_Doppler}
\end{equation}
The term $h^m\left( {{\bm{\tilde \gamma }^m},{\bm{\gamma }}} \right)$ is the measurement function given as
\begin{equation}
{h^m}\left( {{\bm{\tilde \gamma }^m},{\bm{\gamma }}} \right) = \frac{1}{2}{A^m}\Re _\tau ^m\left( {\Delta {\tau ^m}} \right)\Re _f^m\left( {\Delta f^m} \right)\exp \left( {j{\varphi ^m}} \right)
\label{eq:Measurement_equation}
\end{equation}
where ${A^m}$ is the signal amplitude, and ${\varphi^m}$ is the carrier phase. The error terms $\Delta {\tau ^m} = {\tau ^m} - {\tilde \tau ^m}$ and $\Delta f^m = f^m - \tilde f^m$ represent the code-delay and Doppler errors between sample points and the truth.

\subsection{Covariance and gradient properties of correlation model}

The measurement function in (\ref{eq:Measurement_equation}) and the noise covariance in (\ref{eq:Spatial_coherence}) consist of code-delay and Doppler terms, the former characterizes the position-clock bias block ${{\bm{\gamma }}_{\bf{p}}} = {\left[ {\begin{array}{*{20}{c}} {{{\bf{p}}^T}}&{c\delta t} \end{array}} \right]^T}$ and the latter characterizes the velocity-clock drift block ${{\bm{\gamma }}_{\bf{v}}} = {\left[ {\begin{array}{*{20}{c}} {{{\bf{v}}^T}}&{c\delta \dot t} \end{array}} \right]^T}$ \cite{PV decoupling}. Hence, the position-clock bias part is first characterized through the code-delay covariance terms $\Re _\tau ^m\left( {\nabla \tau _{i,j}^m} \right)$ in (\ref{eq:Spatial_coherence}) and measurement term $\partial \Re _\tau ^m\left( {\Delta {\tau ^m}} \right)$ in (\ref{eq:Measurement_equation}).

The analysis starts from the code-delay noise covariance term $\Re _\tau ^m\left( {\nabla \tau _{i,j}^m} \right)$. The code-delay ${\tau ^m}$ of $m$-th satellite is
\begin{equation}
{\tau ^m} = \frac{{{f_c}}}{c}\left( {\left\| {{{\bf{p}}^m_s} - {\bf{p}}} \right\| + c\delta t - c\delta {t^m_s}} \right)
\label{eq:Tao}
\end{equation}
where ${f_c}$ is the PRN code frequency. The subscript ‘$s$’ denotes the satellite-related parameters. For two arbitrary sample points ${{\bm{\tilde \gamma }}^m_{{\bf{p}},i}}$ and ${{\bm{\tilde \gamma }}^m_{{\bf{p}},j}}$, the code-delay difference $\nabla \tau _{i,j}^m$  are related to their position-clock bias difference \cite{Geometric DPE}.
\begin{equation}
\begin{aligned}
&\nabla \tau _{i,j}^m = \tilde \tau _j^m - \tilde \tau _i^m\\
 =& \frac{{{f_c}}}{c}\left( {\left\| {{{\bf{p}}^m_s} - {{{\bf{\tilde p}}}^m_j}} \right\| - \left\| {{{\bf{p}}^m_s} - {{{\bf{\tilde p}}}^m_i}} \right\| + c\delta {{\tilde t}^m_j} - c\delta {{\tilde t}^m_i}} \right)\\
 \approx& \frac{{{f_c}}}{c}\left[ { - {{\left( {{{\bf{u}}^m}} \right)}^T}\nabla {{\bf{p}}^m_{i,j}} + c\nabla \delta {t^m_{i,j}}} \right]\buildrel\textstyle.\over= -\frac{{{f_c}}}{c}{\left( {{{{\bf{\mathord{\buildrel{\lower3pt\hbox{$\scriptscriptstyle\frown$}} 
\over u} }}}^m}} \right)^T}{{\bf{d}}^m_{{\bf p}i,{\bf p}j}}
\end{aligned}
\label{eq:Tao_distance}
\end{equation}
where ${{\bf{d}}^m_{{\bf p}i,{\bf p}j}} = {{\bm{\tilde \gamma }}^m_{{\bf{p}},j}} - {{\bm{\tilde \gamma }}^m_{{\bf{p}},i}} = {\left[ {\begin{array}{*{20}{c}}
{{{\left( {\nabla {{{\bf{ p}}}^m_{i,j}}} \right)}^T}}&{c\nabla \delta {t^m_{i,j}}}
\end{array}} \right]^T}$ represents the position-clock bias difference vector between sample points ${{\bm{\tilde \gamma }}_{{\bf{p}},i}}$ and ${{\bm{\tilde \gamma }}_{{\bf{p}},j}}$.  The dimensionally-extended LOS direction vector is ${{\bf{\mathord{\buildrel{\lower3pt\hbox{$\scriptscriptstyle\frown$}} 
\over u} }}^m} = {\left[ {\begin{array}{*{20}{c}}
{{{\left( {{{\bf{u}}^m}} \right)}^T}}&{ - 1}
\end{array}} \right]^T}$,  where the original LOS direction vector ${{\bf{u}}^m}$ is shown as
\begin{equation}
{{\bf{u}}^m} = \left( {{{\bf{p}}^m_s} - {\bf{p}}} \right)/\left\| {{{\bf{p}}^m_s} - {\bf{p}}} \right\|
\label{eq:LOS_direction}
\end{equation}
The code-delay difference $\nabla \tau _{i,j}^m$ is the projection of the position-clock bias difference ${{\bf{d}}^m_{{\bf p}i,{\bf p}j}}$ onto the extended LOS direction ${{\bf{\mathord{\buildrel{\lower3pt\hbox{$\scriptscriptstyle\frown$}} 
\over u} }}^m}$ as shown in (\ref{eq:Tao_distance}). Substitute (\ref{eq:Tao_distance}) in (\ref{eq:CAF_tao}), the noise covariance term $\Re _\tau ^m\left( {\nabla \tau _{i,j}^m} \right)$ is derived as
\begin{equation}
\begin{aligned}
&\Re _\tau ^m\left( {\nabla \tau _{i,j}^m} \right) = \max \left( {0,1 - \left| {\frac{{{f_c}}}{c}{{\left( {{{{\bf{\mathord{\buildrel{\lower3pt\hbox{$\scriptscriptstyle\frown$}} 
\over u} }}}^m}} \right)}^T}{{\bf{d}}^m_{{\bf p}i,{\bf p}j}}} \right|} \right)\\
 =& \max \left( {0,1 - \frac{{\sqrt 2 {f_c}}}{c}\left\| {{{\bf{d}}^m_{{\bf p}i,{\bf p}j}}} \right\|\left| {\cos \left( {{{\bf{d}}^m_{{\bf p}i,{\bf p}j}},{{{\bf{\mathord{\buildrel{\lower3pt\hbox{$\scriptscriptstyle\frown$}} 
\over u} }}}^m}} \right)} \right|} \right)
\end{aligned}
\label{eq:Tao_coherence}
\end{equation}
From (\ref{eq:Tao_coherence}), for a fixed sample spacing $\|{\bf{d}}^m_{{\bf p}i,{\bf p}j}\|$, the covariance term $\Re_\tau^m(\nabla \tau_{i,j}^m)$ decreases as the absolute projection of ${\bf{d}}^m_{{\bf p}i,{\bf p}j}$ onto the extended LOS direction ${{\bf{\mathord{\buildrel{\lower3pt\hbox{$\scriptscriptstyle\frown$}} 
\over u} }}^m}$ increases. Therefore, an efficient way to reduce $\Re_\tau^m(\nabla \tau_{i,j}^m)$ is to align ${\bf{d}}^m_{{\bf p}i,{\bf p}j}$ with ${{\bf{\mathord{\buildrel{\lower3pt\hbox{$\scriptscriptstyle\frown$}} 
\over u} }}^m}$, i.e., to set $\left| {\cos \left( {{{\bf{d}}^m_{{\bf p}i,{\bf p}j}},{{{\bf{\mathord{\buildrel{\lower3pt\hbox{$\scriptscriptstyle\frown$}} 
\over u} }}}^m}} \right)} \right| = 1$. This motivates placing position-clock bias sample points along the LOS direction.

Next, the code-delay measurement term $\Re _\tau ^m\left( {\Delta {\tau ^m}} \right)$ and its gradient $\partial \Re _\tau ^m\left( {\Delta {\tau ^m}} \right)/\partial {{\bm{\gamma }}_{\bf{p}}}$ with respect to the position-clock bias parameters \(\bm{\gamma}_{\bf p}\) are evaluated
\begin{equation}
\partial \Re _\tau ^m\left( {\Delta {\tau ^m}} \right)/\partial {{\bf{\gamma }}_{\bf{p}}} = \frac{{{f_c}}}{c}{\rm{rect}}\left( {\frac{{\Delta {\tau ^m}}}{2}} \right){\rm{sign}}\left( {\Delta {\tau ^m}} \right){{\bf{\mathord{\buildrel{\lower3pt\hbox{$\scriptscriptstyle\frown$}} 
\over u} }}^m}
\label{eq:Tao_gradient}
\end{equation}
Within the region where the code-delay error satisfies $\left| {\Delta {\tau ^m}} \right| \le 1$, this gradient term reaches its maximum magnitude of $\sqrt{2} f_c/c$. This means that the position-clock bias sample points should be located within the main lobe of the correlation function $\Re _\tau ^m\left( \cdot \right)$; otherwise, no gradient information is available because $\partial \Re _\tau ^m\left( {\Delta {\tau ^m}} \right)/\partial {{\bm{\gamma }}_{\bf{p}}} = \bf 0$.

For the velocity-clock drift part ${{\bm{\gamma }}_{\bf{v}}} = {\left[ {\begin{array}{*{20}{c}} {{{\bf{v}}^T}}&{c\delta \dot t} \end{array}} \right]^T}$, the Doppler of the $m$-th satellite can be written as
\begin{equation}
{f^m} =  - \frac{{{f_L}}}{c}\left[ {{{\left( {{{\bf{u}}^m}} \right)}^T}\left( {{\bf{v}}_s^m - {\bf{v}}} \right) + c\delta \dot t - c\delta \dot t_s^m} \right]
\label{eq:doppler}
\end{equation}
where $f_L$ is the carrier frequency. For two velocity-clock drift sample points, the Doppler difference is given as
\begin{equation}
\nabla f_{i,j}^m 
=  - \frac{{{f_L}}}{c}\left[ { - {{\left( {{{\bf{u}}^m}} \right)}^T}\nabla {\bf{v}}_{i,j}^m + c\nabla \delta \dot t_{i,j}^m} \right] \buildrel\textstyle.\over= \frac{{{f_L}}}{c}{\left( {{{{\bf{\mathord{\buildrel{\lower3pt\hbox{$\scriptscriptstyle\frown$}} 
\over u} }}}^m}} \right)^T}{\bf{d}}_{{\bf{v}}i,{\bf{v}}j}^m
\label{eq:doppler_distance}
\end{equation}
Equation (\ref{eq:doppler_distance}) shows that the Doppler is also determined by the projection of velocity-clock drift sample spacing ${\bf{d}}_{{\bf{v}}i,{\bf{v}}j}^m$ onto the extended LOS direction ${{{{\bf{\mathord{\buildrel{\lower3pt\hbox{$\scriptscriptstyle\frown$}}\over u} }}}^m}}$.

Substituting (\ref{eq:doppler_distance}) into the Doppler noise covariance term $\Re _f^m\left( {\nabla f_{i,j}^m} \right)$ gives
\begin{equation}
\Re _f^m\left( {\nabla f_{i,j}^m} \right) = {\rm{sinc}}\left( {\frac{{{f_L}}}{c}{{\left( {{{{\bf{\mathord{\buildrel{\lower3pt\hbox{$\scriptscriptstyle\frown$}} 
\over u} }}}^m}} \right)}^T}{\bf{d}}_{{\bf{v}}i,{\bf{v}}j}^m{T_c}} \right)
\label{eq:doppler_coherence}
\end{equation}
Within the main lobe of the $\rm sinc$ function, for a fixed spacing $\|{\bf d}^m_{{\bf v}i,{\bf v}j}\|$, aligning the direction of ${\bf d}^m_{{\bf v}i,{\bf v}j}$ with ${{{\bf{\mathord{\buildrel{\lower3pt\hbox{$\scriptscriptstyle\frown$}} 
\over u} }}}^m}$ maximizes the Doppler difference and therefore reduces the Doppler noise covariance $\Re _f^m\left( {\nabla f_{i,j}^m} \right)$. The gradient of the Doppler measurement term with respect to ${{\bm{\gamma }}_{\bf v}}$ is
\begin{equation}
\begin{aligned}
\partial \Re _f^m\left( {\Delta {f^m}} \right)/\partial {{\bm{\gamma }}_{\bf{v}}} = \frac{{{f_L}{T_c}}}{c}{\rm{sinc'}}\left( {\Delta {f^m}{T_c}} \right){{\bf{\mathord{\buildrel{\lower3pt\hbox{$\scriptscriptstyle\frown$}} 
\over u} }}^m}
\end{aligned}
\label{eq:doppler_gradient}
\end{equation}
Within the region $\left| {\Delta {f^m}{T_c}} \right| \le 1$, this gradient term maintains effective sensitivity with a relatively large magnitude. In side lobes of the $\rm sinc$ function, the gradient oscillates and decays to zero, i.e., $\partial \Re _f^m\left( {\Delta {f^m}} \right)/\partial {{\bm{\gamma }}_{\bf{v}}} \to {\bf{0}}$.

In summary, efficient PVT sample points should be placed within the main lobe of the correlation function and aligned with the extended LOS direction 
${{{{\bf{\mathord{\buildrel{\lower3pt\hbox{$\scriptscriptstyle\frown$}}\over u} }}}^m}}$ for each satellite channel. The main-lobe regions are 
$\left| {\Delta {\tau ^m}} \right| \le 1$ for the position-clock bias part and 
$\left| {\Delta {f^m}{T_c}} \right| \le 1$ for the velocity-clock drift part. This strategy maintains effective gradient information and reduces redundancy among samples.

\subsection{The design of LOS sampling strategy}

Following the above summary, the main-lobe constraint and the extended LOS direction jointly define a channel-specific LOS sampling line segment. Let ${{\bm \chi}} \in \left\{ {\bf p},{\bf v} \right\}$ denote the position-clock bias block $({{\bm \chi}}={\bf p})$ and the velocity-clock drift block $({{\bm \chi}}={\bf v})$. The LOS sampling line segment is shown as
\begin{equation}
\ell_{{{\bm \chi}}}^{m}\left(\lambda_{{{\bm \chi}}}\right)
=
{\bm{\gamma}}_{{{\bm \chi}}}
+
\lambda_{{{\bm \chi}}}{{\bf{\mathord{\buildrel{\lower3pt\hbox{$\scriptscriptstyle\frown$}} 
\over u} }}^m},
\qquad
\lambda_{{{\bm \chi}}}\in\left[-\rho_{{{\bm \chi}}},\rho_{{{\bm \chi}}}\right],
\label{eq:Optimal_line_chi}
\end{equation}
where ${{\bf{\mathord{\buildrel{\lower3pt\hbox{$\scriptscriptstyle\frown$}} 
\over u} }}^m}$ is the extended LOS vector of the $m$-th satellite, scalar $\lambda_{{{\bm \chi}}}$ identifies a specific point along the line segment of (\ref{eq:Optimal_line_chi}), and the endpoints $\rho_{{{\bm \chi}}}$ are  
\begin{equation}
\rho_{\bf p} = c/(2f_c); \quad \rho_{\bf v} = c/(2f_LT_c)
\label{eq:rho_chi}
\end{equation}
The range $\left[-\rho_{\bf p},\rho_{\bf p}\right]$  corresponds to the main code-delay lobe $\left|\Delta\tau^m\right|\leq 1$, while the range $\left[-\rho_{\bf v},\rho_{\bf v}\right]$ corresponds to the main Doppler lobe $\left|\Delta f^mT_c\right|\leq 1$.

After the line segment is determined, the LOS-sampled points for $m$-th satellite are obtained by uniform sampling
\begin{equation}
\left\{
{\bm{\tilde\gamma}}_{{\bm \chi},k}^{m}
\right\}_{k=1}^{K_{{\bm \chi}}}
=
\left\{
\ell_{{\bm \chi}}^{m}
\left(\lambda_{{\bm \chi},k}\beta_{{\bm \chi}}\right)
\; \middle| \;
\lambda_{{\bm \chi},k}={ - \frac{{{\alpha _{\bm \chi} }}}{2}, \cdots ,0, \cdots ,\frac{{{\alpha _{\bm \chi} }}}{2}}
\right\}
\label{eq:LOS_sampling_chi}
\end{equation}
where $\alpha_{{\bm \chi}}$ is a positive even integer representing the number of sampling intervals, while $\beta_{{\bm \chi}}$ is the sampling step size, and $K_{{\bm \chi}}=\alpha_{{\bm \chi}}+1$ is the number of samples. Covering the entire main lobe requires $\alpha_{{\bm \chi}}\beta_{{\bm \chi}}=2\rho_{{\bm \chi}}$.
In practice, the true block center ${\bm{\gamma}}_{{\bm \chi}}$ in (\ref{eq:Optimal_line_chi}) is unavailable, and the previous-epoch estimate ${\bm{\hat\gamma}}_{{\bm \chi}}$ is used instead. Thus, the achieved performance also depends on the sampling quality, especially center accuracy and correlation-peak coverage.

After these channel-specific sample points are determined, the estimation criterion for each block ${\bm \chi}$ is formulated as
\begin{equation}
{\bm{\hat\gamma}}_{{\bm \chi}}
=
\mathop{\arg\min}\limits_{{\bm{\gamma}}_{{\bm \chi}}}
\frac{1}{2}
\left(
{\bm{\Re}}_{{\bm \chi}}
-
{\bf h}_{{\bm \chi}}\left({\bm{\gamma}}_{{\bm \chi}}\right)
\right)^{H}
{\bf R}_{{\bm \chi}}^{-1}
\left(
{\bm{\Re}}_{{\bm \chi}}
-
{\bf h}_{{\bm \chi}}\left({\bm{\gamma}}_{{\bm \chi}}\right)
\right)
\label{eq:MLE_criterion}
\end{equation}
Here, matrices ${\bm{\Re}}_{{\bm \chi}}$, ${\bf h}_{{\bm \chi}}\left(\cdot\right)$, and ${\bf R}_{{\bm \chi}}$ collect the correlation values in (\ref{eq:Correlation_model}), measurement model in (\ref{eq:Measurement_equation}), and noise covariance in (\ref{eq:Spatial_coherence}) for block ${\bm \chi}$, respectively. This residual-minimization criterion does not require different satellite channels to share common PVT points, and can be solved by a gradient-based optimizer such as the Levenberg--Marquardt (L-M) method.

\subsection{The information loss and CRB for LOS-sampled DPE}

This subsection quantifies the information loss induced by LOS sampling under different numbers of sampling intervals $\alpha_{{\bm \chi}}$ and sampling step sizes $\beta_{{\bm \chi}}$, and then derives the corresponding CRB for PVT estimation.

The information-loss FIM obtained from the LOS-sampled correlation measurements is defined and can be derived using the properties of LOS-sampled points (\ref{eq:Optimal_line_chi})-(\ref{eq:LOS_sampling_chi}), and properties of corresponding correlation models (\ref{eq:Tao_coherence})-(\ref{eq:Tao_gradient}), (\ref{eq:doppler_coherence})-(\ref{eq:doppler_gradient})
\begin{equation}
{\bf F}_{{\bm \chi}}
=
{\bf J}_{{\bm \chi}}^{H}
{\bf R}_{{\bm \chi}}^{-1}
{\bf J}_{{\bm \chi}}
=
\sum_{m=1}^{M}
\left( {\bf J}_{{\bm \chi}}^{m} \right)^{H}
\left( {\bf R}_{{\bm \chi}}^{m} \right)^{-1}
{\bf J}_{{\bm \chi}}^{m}
\doteq
{\bf G}^{T}
{\bf Q}_{{\bm \chi}}
{\bf G}
\label{eq:F}
\end{equation}
where ${\bf J}_{{\bm \chi}}^{m}$ and ${\bf R}_{{\bm \chi}}^{m}$ are the Jacobian matrix and noise covariance matrix of correlation values for $m$-th satellite, respectively. The first two expressions in (\ref{eq:F}) give the standard FIM definition and its per-satellite contribution. 

The last expression in (\ref{eq:F}) rewrites the FIM into a geometry-weighted form that separates the signal/satellite geometry parameters from the LOS-sampling parameters through
\begin{equation}
{\bf{G}} = {\left[ {\begin{array}{*{20}{c}}
{{{{\bf{\mathord{\buildrel{\lower3pt\hbox{$\scriptscriptstyle\frown$}} 
\over u} }}}^1}}&{{{{\bf{\mathord{\buildrel{\lower3pt\hbox{$\scriptscriptstyle\frown$}} 
\over u} }}}^2}}& \cdots &{{{{\bf{\mathord{\buildrel{\lower3pt\hbox{$\scriptscriptstyle\frown$}} 
\over u} }}}^M}}
\end{array}} \right]^T}
\label{eq:G}
\end{equation}
\begin{equation}
{\bf Q}_{{\bm \chi}}
=
s_{{\bm \chi}}
F_{{\bm \chi}}^{\rm LOS}
\operatorname{diag}
\left(
\left[
\begin{array}{cccc}
C/N_{0}^{1} &
\cdots &
C/N_{0}^{M}
\end{array}
\right]
\right)
\label{eq:Q_chi}
\end{equation}
\begin{equation}
s_{{\bf p}} = f_c^2T_c/c^2; \quad
s_{{\bf v}} = f_L^2T_c^3/c^2
\label{eq:s_chi}
\end{equation}
Here, the geometry matrix ${\bf G}$ contains the extended LOS vectors of all visible satellites, while the weight matrix ${\bf Q}_{{\bm \chi}}$ contains the signal strength, coherent integration time, and the normalized LOS factor. 

The normalized LOS scale factor $F_{{\bm \chi}}^{\rm LOS}$ characterizes the influence of LOS sampling parameters on FIM, derived as
\begin{equation}
F_{{\bm \chi}}^{\rm LOS}
=
\left( {\bf J}_{{\bm \chi}}^{\rm LOS} \right)^{H}
\left( {\bf R}_{{\bm \chi}}^{\rm LOS} \right)^{-1}
{\bf J}_{{\bm \chi}}^{\rm LOS}
\label{eq:F0}
\end{equation}
The normalized covariance matrix ${\bf R}_{{\bm \chi}}^{\rm LOS}$ and normalized Jacobian vector ${\bf J}_{{\bm \chi}}^{\rm LOS}$ are

\begin{equation}
\left[{\bf R}_{{\bm \chi}}^{\rm LOS}\right]_{i,j}
=
\begin{cases}
\max\left(0,1-\frac{2\beta_{\bf p}f_c}{c}\left|j-i\right|\right),
& {\bm \chi}={\bf p}\\[1.0ex]
\operatorname{sinc}\left(\frac{2\beta_{\bf v}f_LT_c}{c}\left|j-i\right|\right),
& {\bm \chi}={\bf v}
\end{cases}
\label{eq:R_LOS_chi}
\end{equation}

\begin{equation}
\left[{\bf J}_{{\bm \chi}}^{\rm LOS}\right]_{k}
=
\begin{cases}
\operatorname{sign}\left(k-\frac{\alpha_{\bf p}}{2}-1\right),
& {\bm \chi}={\bf p}\\[1.0ex]
\operatorname{sinc}^{\prime}\left[
\frac{2\beta_{\bf v}f_LT_c}{c}
\left(k-\frac{\alpha_{\bf v}}{2}-1\right)
\right],
& {\bm \chi}={\bf v}
\end{cases}
\label{eq:J_LOS_chi}
\end{equation}
where $[\cdot]_{i,j}$ denotes the $(i,j)$-th matrix element, $[\cdot]_{k}$ denotes the $k$-th vector element. It can be observed that the normalized LOS factor $F_{{\bm \chi}}^{\rm LOS}$ is determined by LOS sampling parameters $\alpha_{{\bm \chi}}$ and $\beta_{{\bm \chi}}$.

For comparison, the full-information FIM has the same geometry-weighted structure as (\ref{eq:F})-(\ref{eq:s_chi}), but replaces $F_{{\bm \chi}}^{\rm LOS}$ with the corresponding full-information factor \cite{DPE Cramer–Rao Bound}-\cite{DPE CRB analytic}:
\begin{equation}
F_{{\bf p}}^{\rm FULL}
= 8\pi^{2}\left(B_n/f_c\right)^{2}; \quad
F_{{\bf v}}^{\rm FULL}
= \pi^{2}/3
\label{eq:F0_full}
\end{equation}
where $B_n$ is the root-mean-square bandwidth. The information loss percentage of LOS-sampled DPE relative to this full-information benchmark is then quantified by
\begin{equation}
\eta_{{\bm \chi}}
=
\left(
1-
\frac{F_{{\bm \chi}}^{\rm LOS}}
{F_{{\bm \chi}}^{\rm FULL}}
\right)
\times 100\%
\label{eq:F0_loss}
\end{equation}
Finally, the CRB of LOS-sampled DPE is obtained from the inverse of the two decoupled FIM blocks \cite{CRB benchmark}:
\begin{equation}
{\bf CRB}
=
\operatorname{blkdiag}
\left(
{\bf F}_{\bf p}^{-1},
{\bf F}_{\bf v}^{-1}
\right)
\label{eq:CRB}
\end{equation}

In summary, the normalized LOS factor $F_{{\bm \chi}}^{\rm LOS}$ characterizes the information retained by LOS sampling. As the factor $F_{{\bm \chi}}^{\rm LOS}$ approaches the full-information benchmark $F_{{\bm \chi}}^{\rm FULL}$, the information loss percentage $\eta_{{\bm \chi}}$ approaches zero, and the proposed method approaches the full-information CRB with far fewer correlation evaluations.

\section{Performance Analysis}

\subsection{ The information loss and CRB analysis with Monte Carlo simulations} 

\begin{figure}[ht]
    \centering
    \includegraphics[width=1\linewidth]{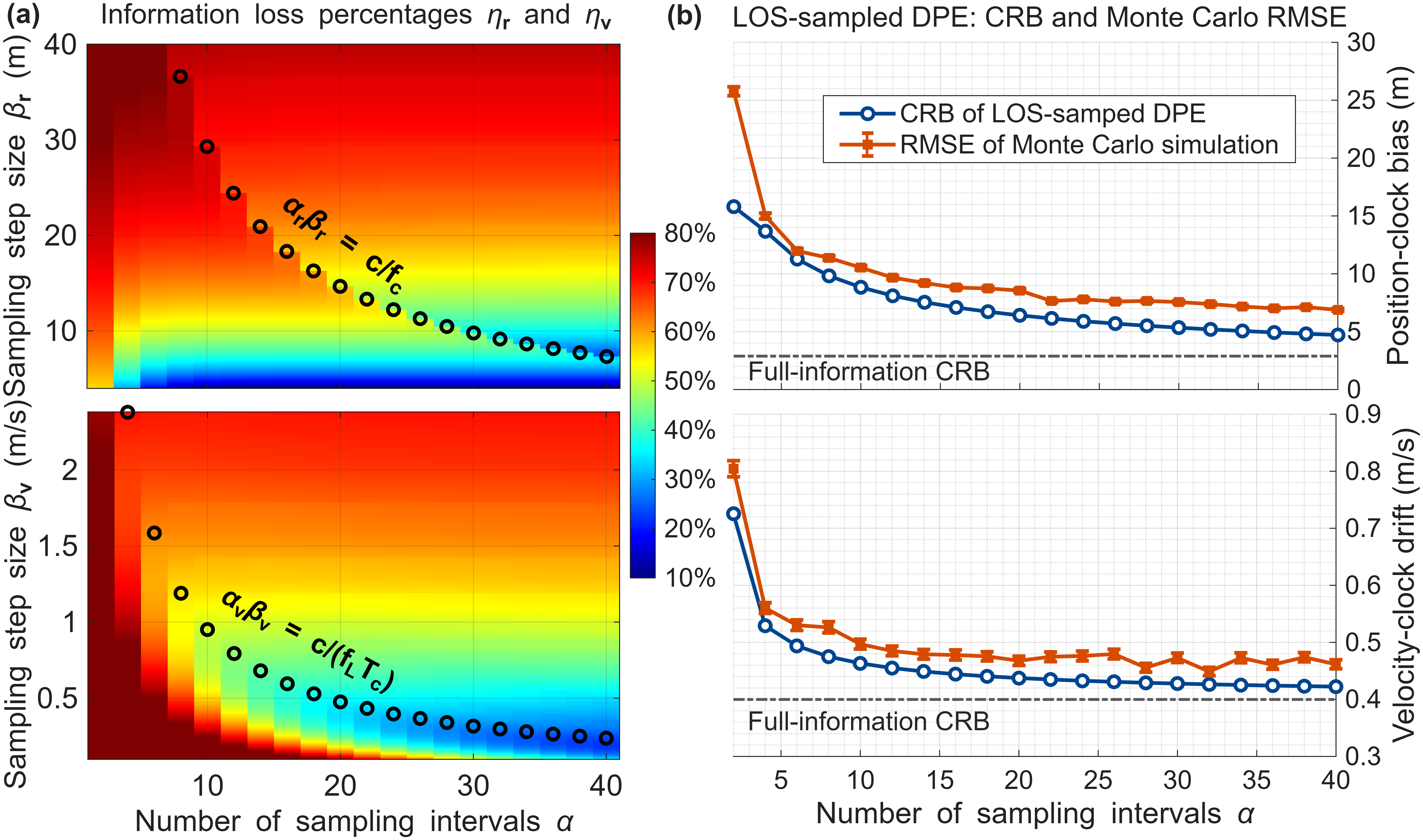}
    \caption{The information loss percentages (a), CRB and Monte Carlo RMSE results (b) for LOS-sampled DPE}
    \label{fig:F0_loss_CRB_RMSE}
\end{figure}

The information loss percentages ${\eta _{{\bf{p}}}}$ and ${\eta _{{\bf{v}}}}$ of LOS-sampled DPE relative to the full-information benchmark are evaluated in Fig. \ref{fig:F0_loss_CRB_RMSE}(a), under different numbers of sampling intervals $\alpha$ and sampling step sizes $\beta$. The sampling parameters that satisfy $\alpha_{\bf p} \beta_{\bf p} = c/{f_c}$ and $\alpha_{\bf v}\beta_{\bf v}=c/(f_L T_c)$ are marked black in Fig. \ref{fig:F0_loss_CRB_RMSE}(a), thereby covering the regions \(\left| {\Delta \tau^m} \right| \le 1\) and $\left| {\Delta {f^m}{T_c}} \right| \le 1$. With \(\alpha_{\bf p} = 40\) and \(\beta_{\bf p} = c/(\alpha_{\bf p} f_c) = 7.33\ \text{m}\) for position-clock bias, and \(\alpha_{\bf v} = 40\) and \(\beta_{\bf v} = c/(\alpha_{\bf v} f_LT_c) = 0.24\ \text{m/s}\) for velocity-clock drift, the LOS sampling incurs approximate information loss percentages of ${\eta _{{\bf{p}}}} \approx 25\% $ and ${\eta _{{\bf{v}}}} \approx 20\% $, respectively.

The theoretical CRB and Monte Carlo RMSE results of LOS-sampled DPE are provided in Fig.~\ref{fig:F0_loss_CRB_RMSE}(b), with the full-information CRB used as the benchmark. The CRB calculations employ the following parameters: $T_c = 20\ \text{ms}$, $C/N_0 = 45\ \text{dB-Hz}$, and $\text{GDOP} = \sqrt{\text{tr}\left((G^TG)^{-1}\right)} = 1.92$. For LOS-sampled DPE, the normalized LOS factors $F_{\bf{p}}^{{\rm{LOS}}}$ and $F_{\bf{v}}^{{\rm{LOS}}}$ corresponds to the curves labeled $\alpha_{\bf p} \beta_{\bf p} = c/{f_c}$ and $\alpha_{\bf v}\beta_{\bf v}=c/(f_L T_c)$ in Fig.~\ref{fig:F0_loss_CRB_RMSE}(a), respectively. In 1000 Monte Carlo trials, PVT estimation is performed using the residual-minimization criterion in (\ref{eq:MLE_criterion}) and the L-M solver. As shown in Fig. \ref{fig:F0_loss_CRB_RMSE}(b), when $\alpha_{\bf p}$ and $\alpha_{\bf v}$ increase, the CRB and Monte Carlo RMSE of LOS-sampled DPE gradually approach the full-information CRB benchmark, consistent with the decreasing information loss percentages ${\eta _{{\bf{p}}}}$ and ${\eta _{{\bf{v}}}}$. Compared with the full-information CRB benchmark, the theoretical position-clock bias CRB of LOS-sampled DPE increases from 2.9 m to about 4.7 m when $\alpha_{\bf p}=40$, while the velocity-clock drift CRB increases from 0.39 m/s to 0.42 m/s when $\alpha_{\bf v}=40$.

\subsection{Experiment results and computational efficiency analysis}

\begin{figure}[ht]
    \centering
    \includegraphics[width=1\linewidth]{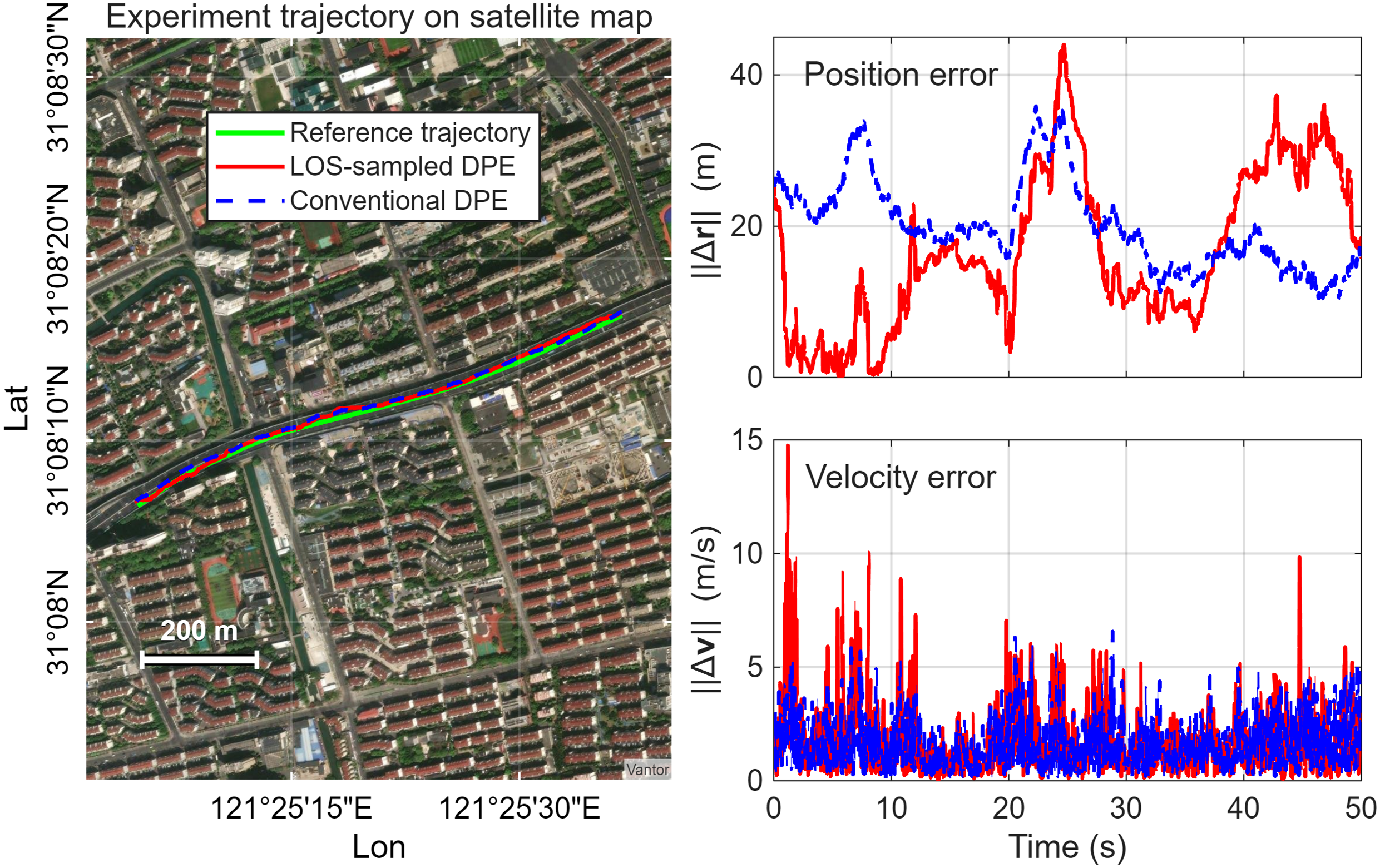}
    \caption{Experimental comparison between LOS-sampled DPE and conventional DPE}
    \label{fig:Experiment_result}
\end{figure}

A vehicle experiment verifies the practical performance of LOS-sampled DPE. The experiment was conducted on the Middle Ring Road in Xuhui District, Shanghai, on September 3, 2022. The IF data were collected by the LabSat 3 Wideband recorder \cite{LabSat}, and the reference trajectory was provided by the Xsens MTi-670 GNSS/INS integrated navigation system \cite{Xsens}. The LOS-sampled DPE is solved by the L-M method, while conventional DPE is solved by grid search on the same IF data. As shown in Fig. \ref{fig:Experiment_result}, both LOS-sampled DPE and conventional DPE achieve comparable PVT estimation accuracy, with about 20 m position error and 4 m/s velocity error.

LOS-sampled DPE reduces the sampling space from four dimensions to one LOS direction per satellite for both position-clock bias and velocity-clock drift blocks. The required number of correlation evaluations is $\left[ {{{\left( {{\alpha _{\bf{p}}} + 1} \right)}^4} + {{\left( {{\alpha _{\bf{v}}} + 1} \right)}^4}} \right]M$ for conventional DPE using grid search, while $\left[ {\left( {{\alpha _{\bf{p}}} + 1} \right) + \left( {{\alpha _{\bf{v}}} + 1} \right)} \right]M$ for LOS-sampled DPE. Thus, the exponential growth of conventional DPE is reduced to linear growth of LOS-sampled DPE. For \(\alpha_{\bf p}=\alpha_{\bf v}=40\) and $M=4$, the number of correlation evaluations decreases from $2.26\times10^7$ to $3.28\times10^2$.

\section{Conclusion}

This paper introduces an efficient LOS-sampled DPE, supported by information-loss CRB analysis, Monte Carlo simulations, and real experiments. In this LOS-sampled DPE, each satellite channel independently retains only PVT sample points aligned with its LOS direction instead of dense common PVT points in the full search space. With proper sampling parameters, LOS-sampled DPE incurs a $20\%$-$25\%$ information loss relative to the full-information benchmark, attains CRBs of 4.7 m and 0.42 m/s, and remains close to the full-information CRBs of 2.9 m and 0.39 m/s, while reducing the number of required correlation values from exponential growth to linear growth. Future work will focus on the robust implementation of LOS-sampled DPE, and combine it with existing DPE acceleration methods to facilitate practical real-time application.

\end{document}